\documentclass[baaa]{baaa}

\usepackage[pdftex]{hyperref}
\usepackage{subfigure}
\usepackage{natbib}
\usepackage{helvet,soul}
\usepackage[font=small]{caption}

\contriblanguage{1}

\contribtype{1}

\thematicarea{8}

\title{Exploring how deviations from the Kerr metric can affect SMBH images}

\titlerunning{How deviations from Kerr metric affect SMBH images}

\author{
F. Agurto-Sepúlveda\inst{1}, J. H. Lagunas\inst{1}, J. Pedreros\inst{1}, B. Bandyopadhyay\inst{1} \& D.R.G Schleicher\inst{1} 
}

\authorrunning{Agurto-Sepúlveda et al.}

\contact{fagurto2016@udec.cl}

\institute{
Departamento de Astronomía, Facultad de Ciencias Físicas y Matemáticas, Universidad de Concepción, Av. Esteban Iturra s/n Barrio Universitario, Casilla 160-C, Concepción, Chile
}

\resumen{Los agujeros negros (BH) son objetos descritos por Relatividad General (GR), en especial por la solución de Kerr. Sin embargo, esta solución sigue estando basada en suposiciones simplificadas. En este marco, exploramos como las desviaciones de la solución de Kerr pueden afectar a las imágenes de los BH supermasivos (SMBH). Esto lo hacemos a través de simulaciones de Transporte de Radiación Relativista General, para una métrica de Kerr-Like. Métrica que posee cuatro funciones no-lineales de desviación libre, estacionaria, axis-simétrica y asintóticamente plana. Para obtener las imágenes, utilizamos el código RAPTOR I, un código abierto de trazado de rayos que admite espacio-tiempos arbitrarios. Y además cuantificamos estas desviaciones a través del estudio de la asimetría, el diámetro y el desplazamiento de la sombra del BH para así compararlas con un BH de Kerr. Nuestros resultados confirman que las desviaciones afectan a la forma de la sombra.  Sin embargo, observamos que la distribución de la materia y las emisividades alrededor del agujero negro también son relevantes, y los resultados no pueden predecirse exclusivamente a partir de la forma del anillo de fotones.}

\abstract{Black holes (BH) are objects described by General Relativity (GR), particularly by the Kerr solution. However, this solution is based on simplifying assumptions. To achieve a more realistic approach, we investigate the impact of deviations from the Kerr solution on the imaging of supermassive black holes (SMBH). We conduct General Relativistic Radiation Transport simulations using a Kerr-Like metric, which includes four nonlinear free deviation functions and is stationary, axis-symmetric, and asymptotically flat. The RAPTOR I code, an open-source ray-tracing code capable of handling arbitrary spacetimes, is employed to generate the images. We analyze the asymmetry, diameter, and shadow displacement of the BH to compare them with those of a Kerr BH. Our findings confirm that these deviations significantly affect the shape of the shadow. However, we acknowledge the importance of considering the matter distribution and emissivities surrounding the BH, as the results cannot be solely inferred from the shape of the photon ring.}


\begin{document}

\maketitle

\section{Introduction}
The first observed BH dates back to the early 1970s \citep{webster1972cygnus}, which was a significant advancement towards proving Einstein's General Relativity. Over the last decades, there has been numerous efforts in detecting these objects. Recently the EHT \citep{collaboration2019first} obtained the first direct image of the SMBH at the center of M87, opening up a new window for studying these objects and testing GR in the strong fields regime. Studies of the strong field tests \citep{gair2013testing,yunes2013gravitational} suggest that astrophysical BH are fully described by Kerr solution but BH don't exist in a perfect vacuum and are not totally axisymmetric but are capable of evolving over time. 

Therefore, there is a need for a more realistic solution in terms of parametric forms of the Kerr metric, or solutions to modified gravity theories. With this context in mind, the Kerr-Like metric \citep{johannsen2013photon} was used for this study, which is a general solution covering both possibilities.
To study the properties of BHs in a Kerr-Like spacetime, one can use GRMHD simulations, which have been extensively studied by authors such as \citet{gammie2003harm,broderick2006radiative, dexter2009fast}, from which it is possible to obtain images of the BH and their shadows using Ray-Tracing algorithms, as developed by \citet{ dexter2016grtrans, bronzwaer2018raptor}. In particular, for this study, the Raptor I code was used, which is capable of using arbitrary spacetimes and also to characterize the properties of these images, the definitions of displacement, diameter, and asymmetry by \citet{johannsen2013photon}. These definitions were used to see how these affect the shadow of BH and possible effects that could be measured in an observation.

\section{Methodology}

\subsection{Kerr-Like metric}
The Kerr-Like metric is a space-time described by \cite{johannsen2013regular} which was constructed with the idea of being regular in the exterior domain, having three constants of motion to be fully integrable and also preserving the separability of the Hamilton-Jacobi equations. It also depends non-linearly on four independent free deviation functions. This metric was initially developed in Boyer-Linquist coordinates

\begin{align}
    g_{tt} &= - \dfrac{\tilde{\Sigma} [\Delta - a^2 A_{2}(r)^{2} \sin^{2}(\theta)]}{[(r^{2} + a^{2})A_{1}(r)- a^{2}A_{2}(r) \sin^{2}(\theta)]^{2}}, \\
    g_{t \phi} &=  - \dfrac{a[(r^{2} + a^{2})A_{1}(r) A_{2}(r) - \Delta]\tilde{\Sigma} \sin^{2}(\theta)}{[(r^{2} + a^{2})A_{1}(r)- a^{2}A_{2}(r) \sin^{2}(\theta)]^{2}}, \\ 
    g_{rr} &=   \dfrac{\tilde{\Sigma}}{\Delta A_{5}(r)}, \\
    g_{\theta \theta} &= \tilde{\Sigma}, \\
    g_{\phi \phi} &=  \dfrac{\tilde{\Sigma} \sin^{2}(\theta) [(r^2 + a^2)^{2} A_{1}(r)^{2} - a^{2}\Delta \sin^{2}(\theta)]}{[(r^{2} + a^{2})A_{1}(r)- a^{2}A_{2}(r) \sin^{2}(\theta)]^{2}},
\end{align}
where $A_{1}(r), A_{2}(r), A_{5}(r)$ are the free functions of the metric and are power series of $M/r$, $\Delta \equiv  r^2 - 2Mr + a^2 $ and $ \tilde{\Sigma} = r^2 + a^2 \cos^{2} \theta + f(r)$ in which $f(r)$ is another free function in terms of power series of $M/r$. Later Johanssen described the same Kerr-Like space-time but in Kerr-Schild coordinates, for more details see \citet{johannsen2013regular}.

\subsection{RAPTOR I}
The Kerr-Like metric was employed in a GRMHD simulation using the open-source RAPTOR I code \citep{bronzwaer2018raptor}. This code, written in C programming language, was designed with two objectives: minimizing physical assumptions in arbitrary space-times and enabling time-dependent radiative transfer. It efficiently utilizes the GPU and CPU of the system. By employing a Ray-tracing algorithm, the code calculates the intensity seen in each pixel of a virtual camera positioned in the observer's frame. The trajectory of photons around the black hole is determined by solving the equation of the null geodesic. Once the photon trajectory values are obtained, the radiative transfer equation is solved to determine the observer intensity \citep{bronzwaer2018raptor}
\begin{align}
    \dfrac{d}{d \overline{\lambda}}\left(\dfrac{I_{\nu,\text{obs}}}{\nu_{\text{obs}}^{3}}\right) =  \dfrac{j_{\nu}}{\nu^{2}}\exp (\tau_{\nu,\text{obs}}(\overline{\lambda})),
\end{align}
where $j_{\nu}$ is the emission coefficient and $ \tau_{\nu,\text{obs}}$ the optical depth at the observer, and $\nu = -k^{\alpha} u_{\alpha}$ the frequency in the plasma frame.

The code uses as initial plasma data file extracted from the outputs of HARM simulations \citep{gammie2003harm} in Kerr-Schild coordinates and we use a synchrotron emission model to calculate the intensity on each pixel of the image.

\subsection{Displacement, diameter, and asymmetry}
In order to quantify the shadows of the BH through the intensity profiles and to observe how much this shadow changes while studying the Kerr-Like spacetime with certain deviation parameters in comparison with the Kerr spacetime, the definition of displacement in the $x$-axis, given by \cite{johannsen2013photon}, is

\begin{align}
    D \equiv \dfrac{|x_{max} - x_{min}|}{2},
    \label{Displacement}
\end{align}

where $x_{max}$ and $x_{min}$, are the locations of the two maximum peaks for the normalized intensity in a horizontal intensity profile. Similarly, it is possible to define an offset in the vertical direction. From here the average radius can be defined as,
\begin{align}
    \langle \overline{R} \rangle \equiv \dfrac{1}{2 \pi} \int_{0}^{2\pi} \overline{R} d\alpha,
    \label{Radius}
\end{align}
where $\overline{R} \equiv \sqrt{(x' - D_{x})^{2} + (y' - D_{y})^{2}}.$

Therefore the average diameter is given by $L \equiv 2 \langle \overline{R} \rangle.$ Thus, the asymmetry of the photon ring is

\begin{align}
    A \equiv 2\sqrt{\dfrac{\int_{0}^{2\pi} \left(\overline{R} - \langle \overline{R} \rangle \right)^{2} d \alpha}{2\pi}}.
    \label{asymmetry}
\end{align}

We compared with the fits for the diameter, displacement, and asymmetry of the BH calculated by \cite{johannsen2013photon} in Kerr and Kerr-Like spacetime, respectively.

\section{Results}
We performed a simulation of an SMBH with the mass of M87* in a Kerr-Like spacetime using a plasma model extracted from the HARM code. Generally, alterations in the spacetime geometry have an impact on the structure of the accretion disk and the observable image of the black hole. Since our main focus is to understand changes due to radiation transport, it is necessary to use the same input concerning the matter distribution and the emissivities (see Table (\ref{tabla1})).

\begin{table}[!t]
\centering
\caption{Setup for the simulation of the SMBH with the distance and mass of M87*, values extracted from \citet{gebhardt2011black,event2019first} }.
\begin{tabular}{lc}
\hline\hline\noalign{\smallskip}
\!\!Variable & \!\!\!\! Value  \\
\hline\noalign{\smallskip}
\!\!Mass  &  $6.2 \times 10^{9} $ $\mathrm{[M_{\odot}]}$\\
\!\!Distance & $16.9 $ $\mathrm{[Mpc]}$\\
\!\!$a$ & $0.9375 \ M$  \\
\!\!$r_{\text{camera}}$ & $10^{4} \ M$\\
\!\!Range for $(\alpha$, $\beta)$ & $[-15,15] \ M$ \\
\!\!Resolution $(x,y)$ & $500$ $\mathrm{[px]}$ \\
\!\!Frequency & $230$ $\mathrm{[GHz]}$ \\
\!\!Inclination (°) & $0,30,60,90$ \\
\!\!$T_p/T_e$ & $1.0$ \\
\!\!$n_e$ & $7.8 \times 10^{5} \ \rho$ $\mathrm{[cm^{-3}]}$ \\
\hline
\end{tabular}
\label{tabla1}
\end{table}

\begin{figure}[!t]
\centering
\includegraphics[width=\columnwidth]{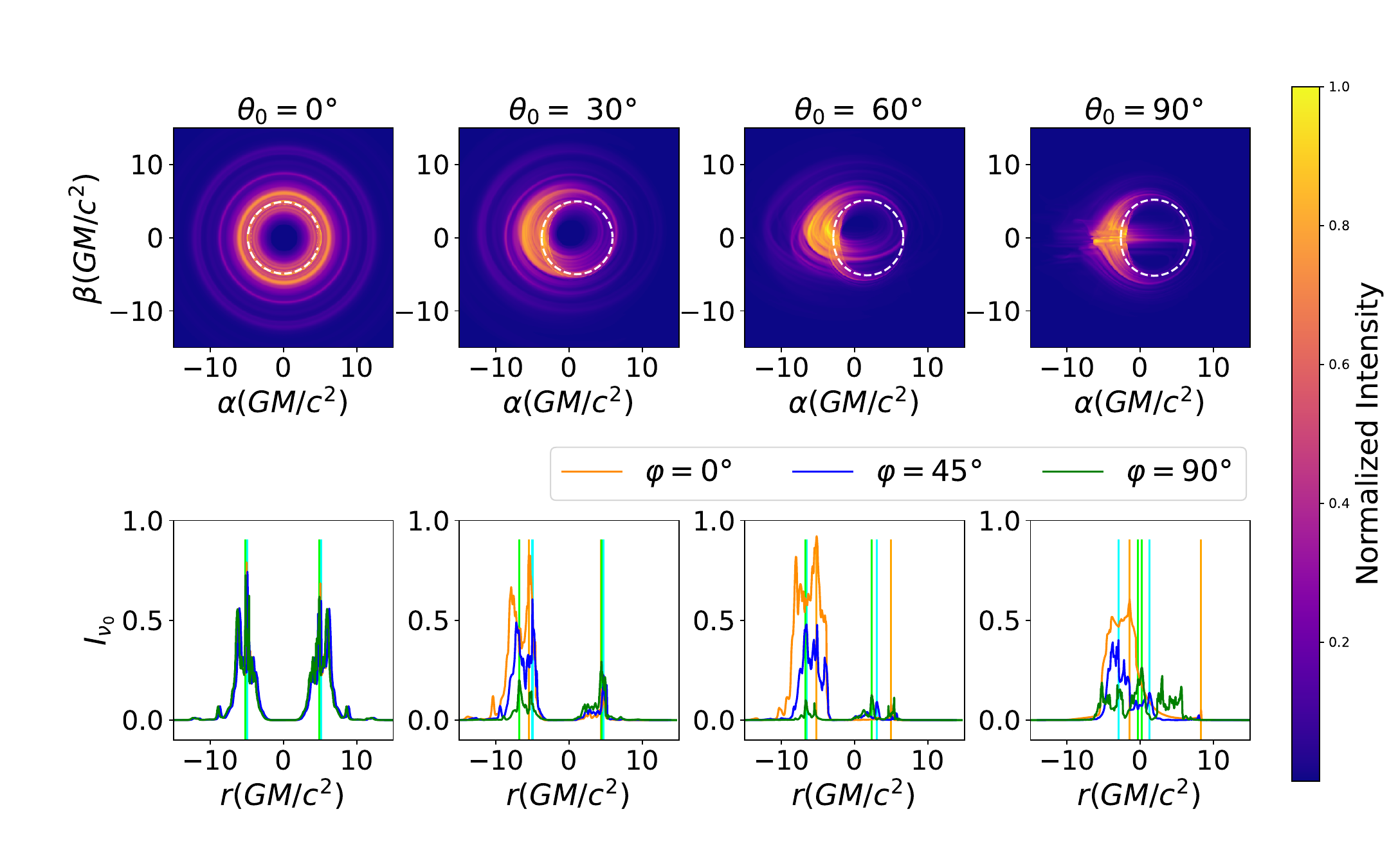}
\caption{Four simulated images of a Kerr-Like BH with the mass of M87* and with deviation parameters $\alpha_{13} = -1$ (\emph{Above images}) with the normalized intensity color bar and below each image the intensity profiles for three different intensity measurement values. }
\label{Figura1}
\end{figure}

Several simulations of this type were conducted with deviation parameters ranging from $\alpha_{13} = [-1, 1]$ and $\alpha_{22} = [-1, 1]$. The most interesting values to observe for these parameters are $\alpha_{13} = -1$ and $\alpha_{22} = 1$, as they exhibit a significant difference in the shadow's diameter, geometry (circularity based on the inclination angle), and displacement from the image center. The upper panel of Fig. \ref{Figura1} displays images from a simulation of a supermassive black hole (SMBH) in a Kerr-like spacetime, captured from various viewing angles spanning from face-on to edge-on, with the mass of $M87*$ being considered. The chosen deviation parameter is $\alpha_{13} = -1$, illustrating the behavior of the BH's shadow by adopting the lower limit for this parameter, with a spin of $a = 0.9375M$. The \textit{white dashed line} represents the circular orbit of the photon ring for a Kerr BH, included for the purpose of comparison with the shadow in our simulation of the Kerr-like metric. Additionally, the lower section displays normalized intensity profiles obtained from data captured at three angles. Vertical lines mark the intensity peaks, allowing for the calculation of $x_{max}$ and $x_{min}$. These values are used to determine the displacement, diameter, and asymmetry of the images. The same analysis is performed for the deviation parameter $\alpha_{22} = 1$.

\begin{figure}[!t]
\centering
\includegraphics[width=\columnwidth]{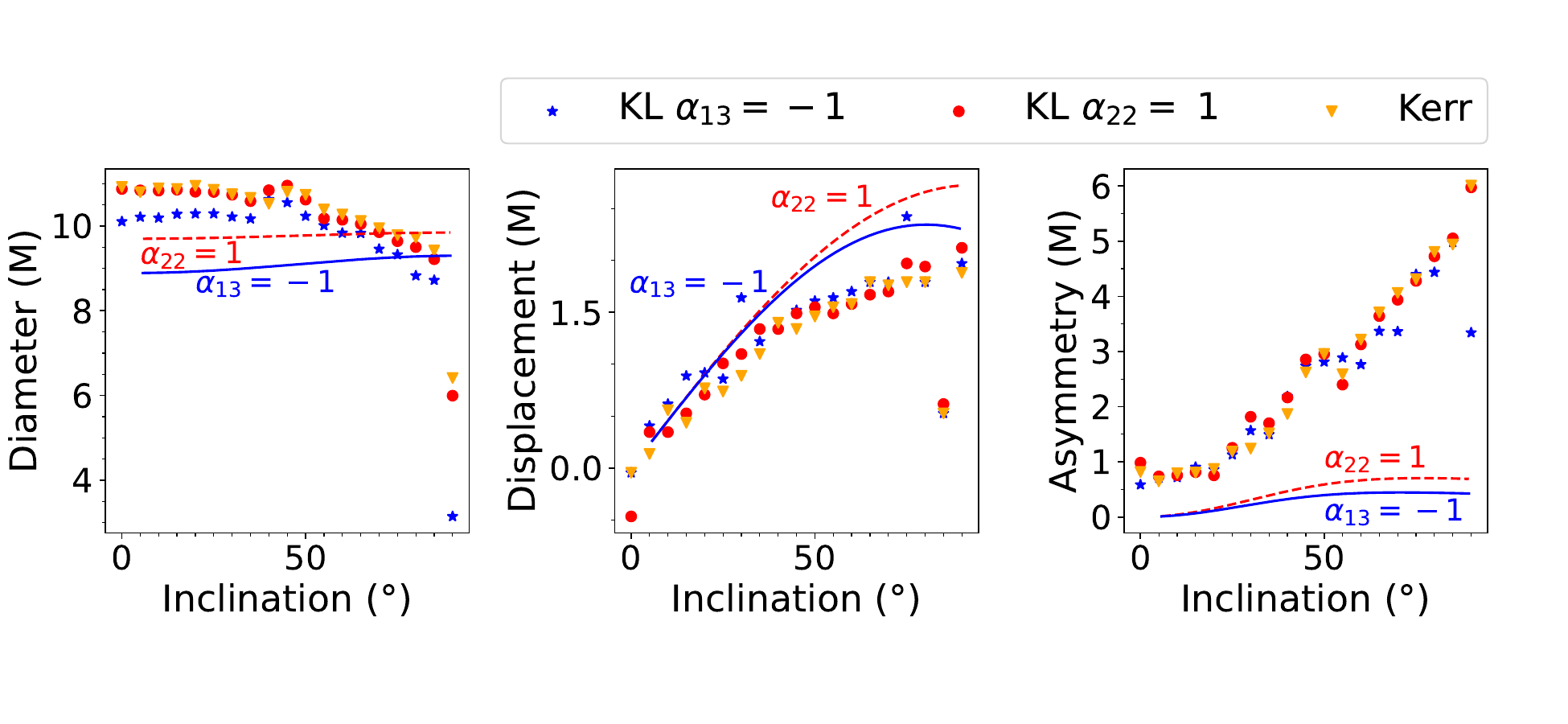}
\caption{Comparison of the diameter, displacement and asymmetry values between the fit calculated by \cite{johannsen2013photon} (\emph{Blue line}) for a Kerr-Like BH with a deviation parameter $\alpha_{13}=-1$ (\emph{Red line}), $\alpha_{22}=1$ (\emph{Blue dashed line}) and a Kerr-Like simulation (\emph{Red dots} and \emph{Blue stars}) of a BH with these deviation parameters, and with the Kerr BH simulation (\emph{Orange triangles}). }
\label{Figura3}
\end{figure}

To characterize and compare these images with the Kerr metric, we utilized equations (\ref{Displacement}-\ref{asymmetry}) and the fit values calculated by \cite{johannsen2013photon}. The comparison is shown in Figure (\ref{Figura3}) for $\alpha_{13} = -1$ and $\alpha_{22} = 1$.

Johanssen's fit, based on the definition of the black hole's shadow edge, differs from our simulations that utilize light sources from the accretion disk.This discrepancy arises from fundamental differences in the parameter definitions compared to our method of intensity profile calculations. However, certain parameters, such as diameter and displacement, closely resemble the expected black hole shadow, facilitating its identification. The inclusion of an accretion disk and plasma also affects the expected fit, as it assumes vacuum conditions for the circular orbit of a test particle. When comparing Kerr simulations with Kerr-Like simulations, their values are more similar. The diameter slightly decreases for both deviations, the displacement is slightly higher than in Kerr, and the asymmetry values generally coincide, except for $\alpha_{13} = -1$ where greater differences are observed.

\section{Conclusions}
This shows, in principle, that the matter accreting around the black hole has a relevant impact on these quantities (diameter, displacement, and asymmetry of the shadow) in comparison with those calculated theoretically by analyzing circular photon orbits. As indicated by \cite{johannsen2013photon} the values of these deviations $\alpha_{13}, \alpha_{22}$ greatly affect the circular orbits of photons and it is possible to notice this type of deviations in the shadow of the hole black compared to the Kerr metric measuring quantities such as asymmetry, diameter, and shadow displacement. 

\begin{acknowledgement}
\texttt{We thanks to Julio Oliva for enlightening comments. We gratefully acknowledge support by the ANID BASAL projects ACE210002 and FB210003, as well as via the Millenium Nucleus NCN19-058 (TITANs).}
\end{acknowledgement}


\bibliographystyle{baaa}
\small
\bibliography{bibliografia}
 
\end{document}